\documentclass[reprint,amsmath,amssymb,aps]{revtex4-2}
\usepackage[colorlinks,citecolor=blue,linkcolor=blue,anchorcolor=blue,filecolor=blue,
urlcolor=blue]{hyperref}
\usepackage{graphicx}
\usepackage{ulem}
\usepackage{dcolumn}
\usepackage{bm}
\usepackage{todonotes}

\begin{document}

\title{Quark confinement in an equiparticle quark model: application to stellar matter}
\author{Isabella Marzola$^1$, S\'ergio B. Duarte$^2$, C\'esar H. Lenzi$^1$, and Odilon Louren\c{c}o$^{1,3}$}
\affiliation{
\mbox{$^1$Departamento de F\'isica, Instituto Tecnol\'ogico de Aeron\'autica, DCTA, 12228-900, S\~ao Jos\'e dos Campos, SP, Brazil}
\\
\mbox{$^2$Centro Brasileiro de Pesquisas F\'isicas, 22290-180, Urca, Rio de Janeiro, Brazil}
\\
\mbox{$^3$Universit\'e de Lyon, Universit\'e Claude Bernard Lyon 1, CNRS/IN2P3, IP2I Lyon, UMR 5822, F-69622, Villeurbanne, France}
}

\date{\today}

\begin{abstract} 
We perform an improvement in a thermodynamical consistent model with density dependent quark masses ($m'_{u,d,s}$) by introducing effects of quark confinement/deconfinement phase transition, at high density regime and zero temperature, by means of the traced Polyakov loop ($\Phi$). We use realistic values for the current quark masses, provided by the Particle Data Group, and replace the constants of the interacting part of $m'_{u,d,s}$ by functions of $\Phi$, leading to a first order phase transition structure, for symmetric and stellar quark matter, with $\Phi$ being the order parameter. We show that the improved model points out the direction of the chiral symmetry restoration due to the emergence of a deconfined phase. In another application, we construct quark stars mass-radius profiles, obtained from this new model, and show to be possible to satisfy recent astrophysical observational data coming from the LIGO and Virgo Collaboration, and the NICER mission concerning the millisecond pulsars PSR J0030+0451, and PSR J0740+6620.
\end{abstract}

\maketitle

\section{Introduction}

The quantum chromodynamics (QCD)~\citep{qcd1,qcd2,qcd3} theory establishes that quarks and gluons are, at the most fundamental level, the degrees of freedom of systems composed by strongly interacting particles. In principle, infinite nuclear matter, finite nuclei, or even stellar matter should also be described directly by this theory. However, this is not a trivial task to be implemented due to the nonperturbative nature of the QCD infrared region. Because of that, different approaches are used to treat systems of quarks and gluons, such as the lattice calculations~\citep{lattice1,lattice2}, based on sophisticated numerical simulations, application of Dyson-Schwinger equations in Euclidean space~\citep{dse1,dse2}, and the effective/phenomenological quark/gluons models, developed in order to present as many similarities with QCD as possible. In that direction, many options were constructed over the years as, for instance, the Massachusetts Institute of Technology (MIT) bag model~\citep{mit1,mit2}, in which the building block particles are submitted to a confining potential mathematically represented by a ``bag'' constant, and the Nambu-Jona-Lasino (NJL) model~\citep{njl1,njl2}, in which  dynamical breaking of chiral symmetry is taken into account. Another class of effective models explicitly considers density and/or temperature dependent quark masses~\citep{ddmasses1,ddmasses2} (in the MIT bag model these quantities are kept fixed, and in the NJL model such dependencies are obtained in an implicit way). 

A very relevant issue verified in these density dependent quark masses models is the break of thermodynamic consistency observed in their equations of state. Basically, this violation emerges because the pressure at a density corresponding to the minimum of the energy per baryon is not vanishing as it should be. However, such a problem was solved in~\cite{Xia:2014}, a paper in which the authors proposed a suitable expression for the density dependent equivalent quark masses, along with the direct connection between the quark Fermi momentum, at $T=0$ regime, with an effective chemical potential instead of the real one. The resulting model was named as the equiparticle~(EQP) model, in which both, energy density and particle number densities, have the same form of the respective free particle system quantities. The concept of effective chemical potentials used in that description is also useful at finite temperature regime~\citep{Xia:2014}.

Since the thermodynamic inconsistency has been circumvented, the authors of~\cite{Xia:2014} were able to apply the EQP model to the description of strange quark stars. The concept of the existence of strange matter as the true ground state instead of hadrons was proposed by Bodmer~\citep{Bodmer:1971}, who claimed that quark matter could have lower energy per baryon than normal nucleus~($^{56}\rm Fe$), and Witten~\citep{Witten:1984}, who considered stable strange matter as composed by quarks up, down and strange, and also proposed the existence of strange quark stars. These ideas were named as the Bodmer-Witten conjecture/hypothesis, and a final experimental/observational probe is not available yet. Nevertheless, in the last years, a huge number of new astrophysical data has been arisen, specially due to the recent detection, made by the LIGO and Virgo Collaboration~\citep{Abbott_2017,Abbott_2018,Abbott_2020-2}, of gravitational waves coming from a binary system with its respective electromagnetic counterparts also detected by many observatories~\citep{ligo-em}, and also due to measurements from the x-ray telescope installed on the International Space Station, named as the NASA's Neutron star Interior Composition Explorer (NICER), regarding the massive millisecond pulsars PSR~J0030+0451~\citep{Riley_2019,Miller_2019} and PSR~J0740+6620~\citep{Riley_2021,Miller_2021}. These new data has been often used as a constraint to test many effective hadronic~\citep{defrmf,defsky,lucas} and quark models nowadays. 

In this paper, we propose an improvement in the EQP model by including on it the confinement/deconfinement phase transition phenomenology (PTP), expected to occur in strong interaction systems at high density regions and at zero temperature regime. We follow the procedure performed in~\citep{Mattos:2019,Mattos:2021,Mattos:prd} and describe the PTP through the inclusion of the traced Polyakov loop~($\Phi$) by making the free parameters of the model suitable functions of $\Phi$. We show that the new improved model exhibits a first-order phase transition structure that can be properly identified through the analysis of its order parameter $\Phi$, or through the signatures presented in the grand-canonical thermodynamical potential as a function of both, $\Phi$, and the chemical potential of the system. The new model is shown to be in the direction of chiral symmetry restoration since a reduction in the values of the quark masses is now verified. We also investigate the capability of the model in producing quark stars mass-radius diagrams compatible with the aforementioned astrophysical observational data. Furthermore, it is verified that a branch of such diagrams is constructed from equations of state representing the deconfined quark phase. For the sake of completeness, we emphasize here that quark stars and hybrid stars represent only two of the possibilities to explain the observed compact objects. There are many equations of state for neutron stars that are also consistent with the same observational data, see for instance~\cite{pearson} and references therein.

The study described above is structured in the paper as follows. In Sec.~\ref{section:formalism} we show the main quantities related to the original equiparticle model. We also obtain the stability window by using recent values of the current quark masses provided by the Particle Data Group~(PDG)~\citep{Workman:2022ynf}. In Sec.~\ref{section:peqpm} we discuss the inclusion of the confinement effects in the model by introducing $\Phi$ in the density dependent quark masses. We investigate the modifications generated by this phenomenology in both, symmetric and stellar quark matter. For the latter, we also construct strange quark stars mass-radius profiles for different parametrizations of the improved model, named as Polyakov-Equiparticle model, and show they are compatible with the astrophysical constraints mentioned before. All study presented here is performed at zero temperature regime.

\section{Equiparticle model}
\label{section:formalism}

\subsection{Main thermodynamical quantities}

We start by presenting the main thermodynamical quantities derived in~\cite{Xia:2014} and used in~\cite{Backes:2020}, where a baryonic density~($\rho_{b}$) dependent model at zero temperature~($T$) is proposed, namely, the EQP model. Since the study done in this work focus on the equations at zero temperature, we restrict ourselves to show the formalism concerning this regime. We assume a system composed of up~($u$), down~($d$) and strange ($s$) quarks with masses being  
\begin{equation}
m_{i} = m_{i0} + m_I = m_{i0} + \frac{D}{\rho_b^{1/3}} + C\rho_{b}^{1/3},
\label{eq:masses}
\end{equation}
where $m_{i0}$ ($i$ = $u$, $d$, $s$) is the current mass of the $i$ quark and $m_I$ is its density dependent part. The parameters $C$ and $D$ are responsible for the interaction effects between quarks and are chosen by following the same criteria of the stability window used in~\cite{Backes:2020}, which will be better explained in the next subsection. It is important to mention that the parameter $C$ is responsible for achieving higher stellar masses when the model is applied to pure quark or hybrid stars, which reinforces the need to include the parameter in the mass scaling obtained in~\cite{Xia:2014} and presented in Eq.~\eqref{eq:masses}.

In order to ensure thermodynamic consistency, the concept of effective chemical potentials, explained in~\cite{Peng:2008}, is incorporated into the model through the Fermi momentum given by
\begin{equation}
k_{Fi} = \sqrt{\mu_i^{*2} - m_i^2},
   \label{eq:fermi-momentum}
\end{equation}
where $\mu_i^{*}$ is the effective chemical potential of the $i$ quark. The relationship between the effective and real chemical potentials reads
\begin{equation}
    \mu_i = \mu_i^* + \frac{1}{3}\frac{\partial m_I}{\partial \rho_b} \frac{\partial \Omega_0}{\partial m_I}.
    \label{eq:real-effective-chemical-potentials}
\end{equation}

Furthermore, the quark density is connected to the Fermi momentum through the following equation
\begin{equation}
    \rho_i = \frac{\gamma k_{Fi}^{3}}{6\pi^2},
    \label{eq:particle-density}
\end{equation}
with $\gamma$ being the degeneracy factor = 3 (color) x 2 (spin). The quark density relates to the baryonic one through
\begin{equation}
    \rho_{b} = \frac{1}{3} \sum_i \rho_i.
    \label{eq:baryon-number-density}
\end{equation}

Energy density and pressure are given, respectively by
\begin{equation}
    \epsilon = \Omega_0 - \sum_i \mu_i^* \frac{\partial\Omega_0}{\partial\mu_i^*},
    \label{eq:energy-density}
\end{equation}
and
\begin{equation}
\begin{split}
    P &= - \Omega_0 + \sum_{i,j} \frac{\partial\Omega_0}{\partial m_j} \rho_i \frac{\partial m_j}{\partial \rho_i} 
    = -\Omega_{0}+\rho_{\mathrm{b}} \frac{\partial m_{\mathrm{I}}}{\partial \rho_{\mathrm{b}}} \frac{\partial \Omega_{0}}{\partial m_{\mathrm{I}}}.
    \label{eq:pressure}
\end{split}
\end{equation}
The quantity identified as $\Omega_0$ refers to the particle contribution of a free system, which for the unpaired case is given by~\cite{Peng:2000}
\begin{eqnarray}
&\Omega_0&= \nonumber \\
&-&\sum_i \frac{\gamma}{24 \pi^2} \left[ \mu_i^*k_{Fi} \left( k_{Fi}^2 - \frac{3}{2}m_i^2 \right) + \frac{3}{2} m_i^4 \ln \frac{\mu_i^* + k_{Fi}}{m_i} \right],\nonumber\\
\label{eq:thermodynamic-potential-free-system}
\end{eqnarray}
allowing us to present explicitly the derivatives in Eq.~\eqref{eq:pressure}. The first one reads
\begin{equation}
\begin{split}
   \frac{\partial m_I}{\partial \rho_b} = -\frac{D}{3\rho_b^{4/3}} + \frac{C}{3\rho_b^{2/3}},
\end{split}
\end{equation}
and the second as
\begin{equation}
\begin{split}
    \frac{\partial \Omega_0}{\partial m_I} &  =  \sum_i \frac{\gamma m_i}{4 \pi^2} \left[  \mu_i^*k_{Fi} - m_i^2 \mbox{ln}\frac{ \mu_i^* + k_{Fi}}{m_i} \right].
    \label{derivmcont}
\end{split}
\end{equation}

The grand-canonical thermodynamical potential can be obtained by taking $\Omega = -P$. Finally, an explicit form for the energy density is obtained from the direct evaluation of the derivative in Eq.~\eqref{eq:energy-density}, or alternatively from the sum of the $i$ quark energy densities ($\epsilon = \sum_i \epsilon_i$), where $\epsilon_i$ is given in~\cite{Wen:2005} as 
\begin{equation}
\begin{split}
 \epsilon_i &=\frac{\gamma}{2\pi^2} \int_{0}^{k_{Fi}} k^{2}\sqrt{k^{2} + m_{i}^{2}}dk.
 \label{eq:i-energy-density-integral}
\end{split}
\end{equation}

Since the model is carefully treated in terms of thermodynamic consistency, all equations above are according to the fundamental thermodynamics, from which the real chemical potentials are obtained, for example. One can prove that by taking the derivative of $-\Omega_0$ with respect to $\mu_i^*$ and comparing it to the particle number density in Eq.\eqref{eq:particle-density} to verify that they are equivalent, leading us to conclude that this relation is consistent with the ones from fundamental thermodynamics. In other words: 
\begin{equation}
\rho_i = -\frac{\partial\Omega_0}{\partial\mu_i^*}  =
\frac{\partial P}{\partial\mu_i}.
\label{eq:consist-verification}
\end{equation}

Another important verification can be done for the energy density through the following calculation:
\begin{equation}
\begin{split}
 P + \epsilon &= \mu\rho \\
 \sum_{i} P_{i} + \sum_{i} \epsilon_{i} &= \sum_{i}\mu_{i}\rho_{i}.
 \label{eq:consist-verification2}
\end{split}
\end{equation}
It is also worth to mention that even with different quark mass scaling being applied, such as the ones from~\cite{Fowler:1981,Chakra:1991,Peng:2000,Xiaoping:2004}, all equations remain thermodynamically consistent. 

\subsection{Symmetric and stellar matter}
\label{symstellarmatter}

For symmetric quark matter, the system must obey the condition of $\mu_{u} = \mu_{d} = \mu_{s} \equiv \mu$. Since the model introduces effective chemical potentials in order to maintain thermodynamic consistency, it is possible to find a equivalent relationship between these quantities instead. From Eq.~\eqref{eq:real-effective-chemical-potentials} we see that the latter term is what differs $\mu^{*}_{i}$ from $\mu_{i}$ and, by being the same for every quark flavor, it allows us to find $\mu^{*}_{u} = \mu^{*}_{d} = \mu^{*}_{s}$. It is worth to notice that the definition of symmetric matter done here considers only equal chemical potentials between quarks, which means that the quarks masses, and densities, are not the same. This concept is different from the one described in hadronic models, where nucleons have equal densities and equal chemical potentials, since their rest masses are almost the same. Here, all quarks masses has its own value, since we are relying on the values provided by PDG~\citep{Workman:2022ynf} to model as close as possible the reality of what symmetric matter could be, as well as stellar matter. Therefore, although quarks $u$ and $d$ have low and relatively close masses, the strange quark has a higher one. Therefore, the quarks densities will be different from each other.

For the stellar matter at zero temperature and high density regime, effective quark models are useful in order to describe compact stars, such as pure quark or hybrid ones~\citep{Schertler:1999, Ranea:2019, Hanauske:2001, Menezes:2006, Lenzi:2010, Drago:2020, Pereira:2016, Ferreira:2020}. The characteristics found in these systems are charge neutrality and beta equilibrium conditions. Due to the presence of leptons (electrons in this case), weak interactions happen such as $d, s \leftrightarrow u + e + \bar\nu_e$. After the Urca process \citep{Lattimer:1991,YAKOVLEV:20011}, where the compact starts cooldown by emitting neutrinos, and consequently, losing a lot of its initial energy, beta equilibrium takes place once neutrinos have left the system, and its condition is expressed in terms of the chemical potentials of the particles, namely, $\mu_u + \mu_e = \mu_d = \mu_s$. As well as done previously for symmetric matter, this relationship can be given in terms of the effective chemical potentials as $\mu_u^* + \mu_e = \mu_d^* = \mu_s^*$, where we see that the electron chemical potential remains the same because electrons have constant mass and do not participate in strong interactions. Furthermore, since compact stars are electrically neutral objects, the charge neutrality condition requires that $\frac{2}{3}\rho_u - \frac{1}{3}\rho_d - \frac{1}{3}\rho_s - \rho_e = 0$. Finally, total energy density ($\epsilon$) and pressure ($p$) of matter in beta equilibrium take into account the leptons contribution. In the case in which only (massless) electrons are included, these expressions are given by  
\begin{eqnarray}
\varepsilon &=& \epsilon + \frac{\mu_e^4}{4\pi^2}
\label{det}
\end{eqnarray}
and
\begin{eqnarray} 
p = P + \frac{\mu_e^4}{12\pi^2},
\label{presst}
\end{eqnarray}
respectively, with $\mu_e=(3\pi^2\rho_e)^{1/3}$.

\subsection{Stability window for recent current quark masses}
\label{stabwindow}

The study of strange quark matter (SQM) done through phenomenological effective models takes Bodmer-Witten hypothesis~\citep{Bodmer:1971,Witten:1984} into account. The stability of SQM is established by investigating it against nuclear matter and is represented by the stability window. In order to construct it from the EQP model, we take a large set of values for the $C$ and $\sqrt{D}$ parameters. Then, the minimum of the energy per baryon, namely, $(E/A)_{\mbox{\tiny min}}=(\mathcal{E}/\rho_b)_{\mbox{\tiny min}}$ where $\mathcal{E}=\epsilon$ ($\varepsilon$) for symmetric (stellar) matter, is evaluated and classified according to the following criteria:

(i) SQM stable: $(E/A)_{\mbox{\tiny min}}\leqslant 930$~MeV. The minimum energy per baryon is lower than the binding energy of $^{56}$Fe.

(ii) SQM metastable: $930\mbox{ MeV}<(E/A)_{\mbox{\tiny min}}\leqslant 939$~MeV (nucleon mass).

(iii) SQM unstable: $(E/A)_{\mbox{\tiny min}}> 939$~MeV.

Furthermore, the two-flavor quark matter (2QM) must be unstable when applied in the model for the same set of $C$ and $\sqrt{D}$ parameters, because we did not find matter containing deconfined $u$ and $d$ quarks in nature neither in terrestrial experiments, i.e., the condition of $(E/A)_{\mbox{\tiny min}}> 930$~MeV has to be satisfied in this case.

\begin{figure}
\includegraphics[scale=0.30]{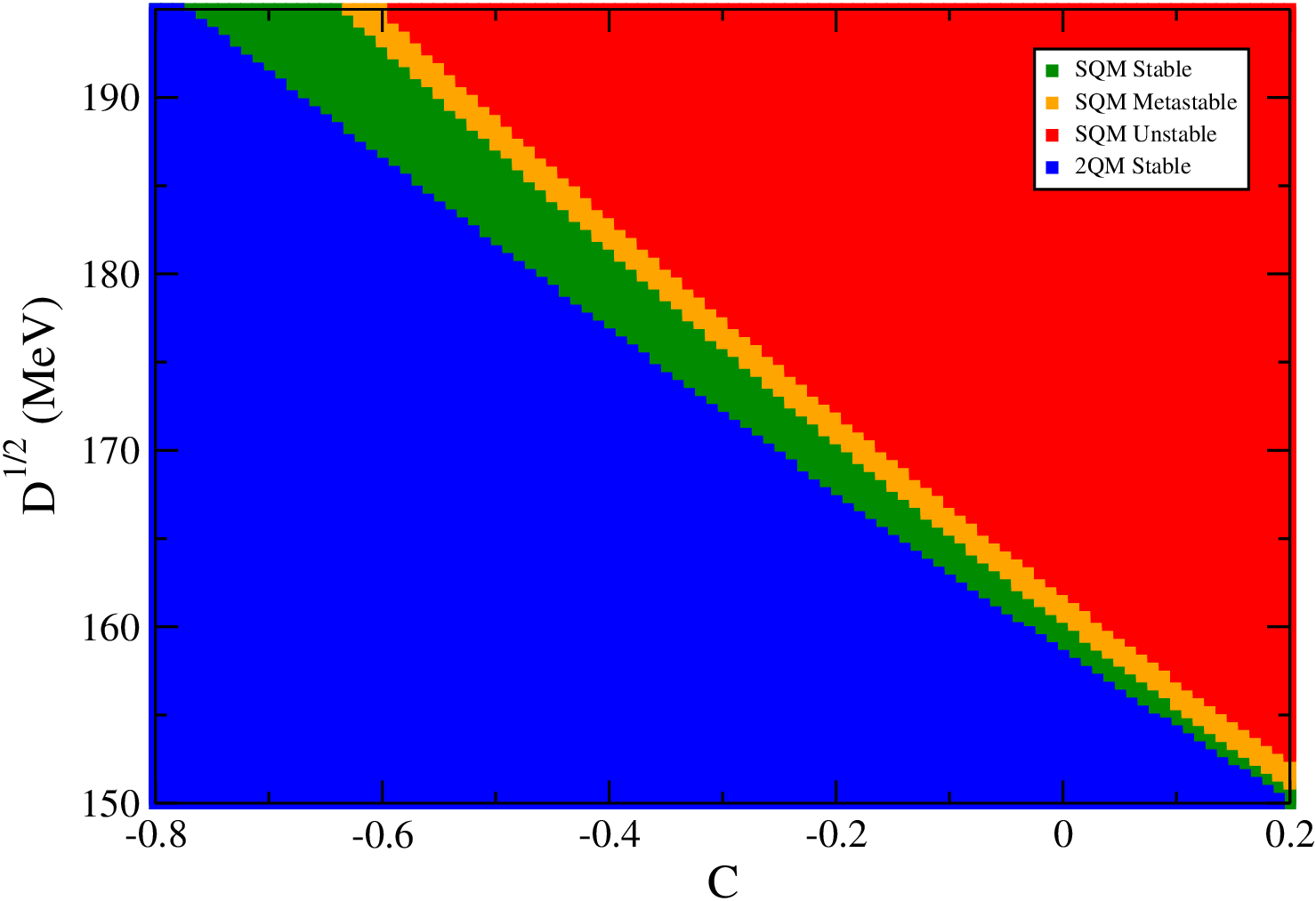}
\caption{Stability window for symmetric matter: EQP model.}
\label{stability-window-symmetric-matter}
\end{figure}

The first results of our study refer to the calculation of the stability window for both, symmetric and stellar matter. Differently from the analysis performed in~\cite{Backes:2020}, here we consider the recent data published by PDG~\citep{Workman:2022ynf} regarding the ranges for the current quark masses, namely, $m_{u0} = 2.16^{+0.49}_{-0.26}$, $m_{d0} = 4.67^{+0.48}_{-0.17}$ MeV and $m_{s0} = 93.4^{+8.6}_{-3.4}$ MeV. For symmetric~(stellar) matter we use $m_{u0} = 1.90$~MeV~($2.16$~MeV), $m_{d0} = 4.67$~MeV~($5.15$~MeV), and $m_{s0} = 93.4$~MeV~($90$~MeV). In Fig.~\ref{stability-window-symmetric-matter} we show the stability window for symmetric matter.
The area of interest of our study corresponds to the green one, where SQM is absolutely stable. The blue region below this area corresponds to where the 2QM is stable, which means that this is a region where the SQM is forbidden to happen. Above the SQM stable area, we have the metastability region in orange and the unstable SQM in red.

\begin{figure}
\includegraphics[scale=0.30]{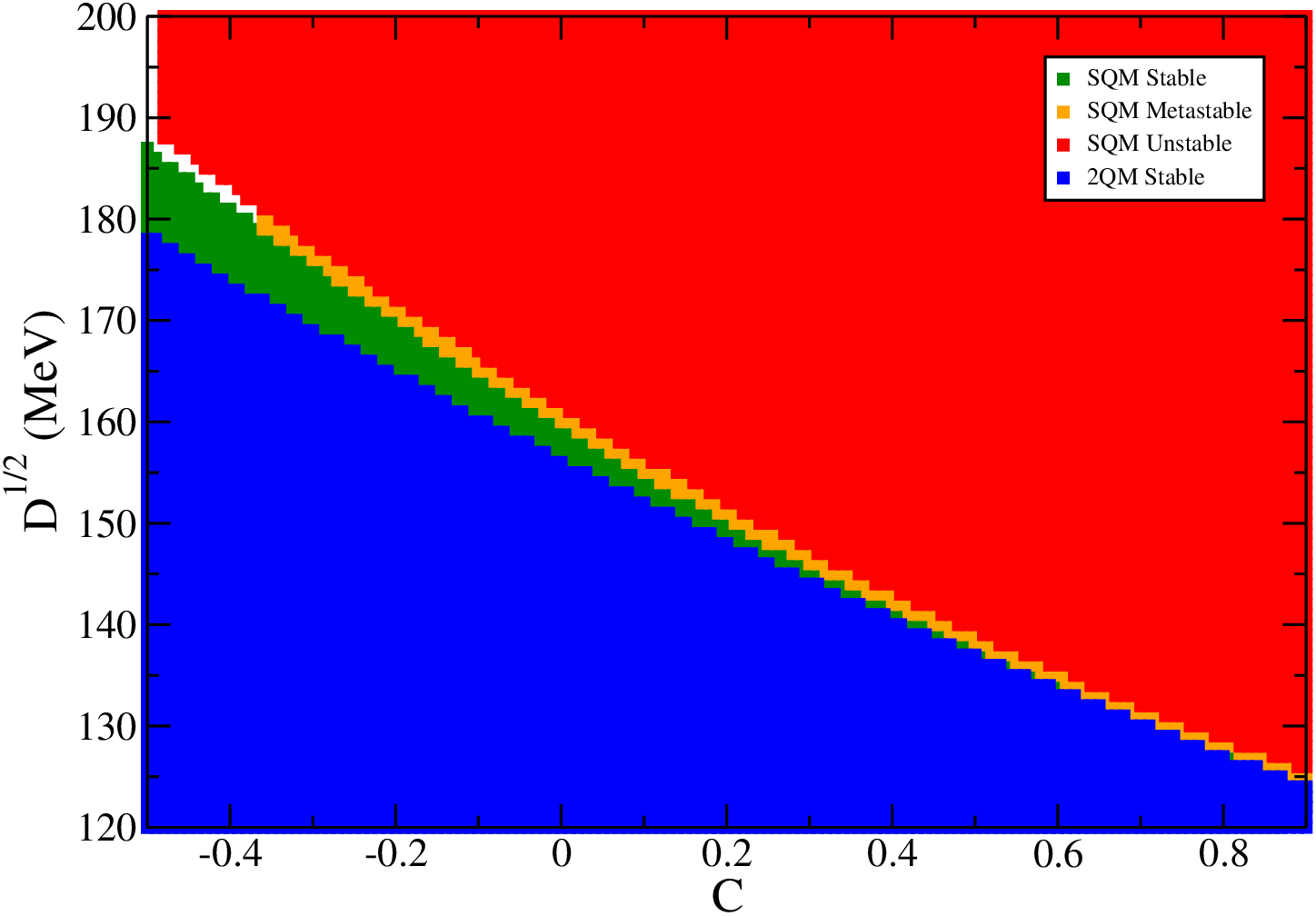}
\caption{Stability window for stellar matter: EQP model.}
\label{stability-window-stellar-matter}
\end{figure}

For stellar matter, we have again the lower region as the forbidden one, where 2QM would be stable, as shown in blue in Fig.~\ref{stability-window-stellar-matter}. 
The SQM stable region is indicated by green and the metastable one is within the orange region. In red, once more, there is the unstable region. Since we have both stability windows, it is possible to see that the first one (symmetric matter, Fig.~\ref{stability-window-symmetric-matter}) includes higher values of $\sqrt{D}$ and lower values of $C$ in comparison with what occurs in the second window (stellar matter, Fig.~\ref{stability-window-stellar-matter}). On the other hand, the stellar stability window contemplates higher values for $C$, which are responsible for providing higher values of maximum masses when it is applied to pure quark or hybrid stars. Besides, the stable and metastable region of stellar matter is thinner than the one of symmetric matter. These differences make it clear that since electrons are entered into the system, they make a huge impact on the model results.

\section{Polyakov equiparticle model}
\label{section:peqpm}

The purpose of our work is to introduce the confinement/deconfinement phase transition in the quark system described by the EQP model. This phenomenology is implemented in an effective way through the inclusion of the traced Polyakov loop, 
\begin{equation}
\begin{split}
\Phi &= \frac{1}{3}Tr \left[ exp\left( i \int_{0}^{1/T}d\tau A_{4} \right) \right],\\
     &= \frac{1}{3}[e^{i(\phi_{3} + \phi_{8}/\sqrt{3})} + e^{i(- \phi_{3} + \phi_{8}/\sqrt{3})} + e^{- 2i\phi_{8}/\sqrt{3}}],
\label{eq:polyakov-loop}
\end{split}
\end{equation}
where, $A_4 = iA_{0} \equiv T\phi$, $A_{0} = gA_{a}^{0}\lambda_{a}/2$ (g is the gauge coupling) and $\phi = \phi_{3}\lambda_{3} + \phi_{8}\lambda_{8}$ (Polyakov gauge). This quantity mimics the gluonic dynamics of the strong interaction and was firstly used in the NJL model~\citep{njl1,njl2,njl3,njl4,njl5,njl6} at finite temperature in~\cite{Fukushima:2004} to generate the so-called Polyakov-Nambu-Jona-Lasinio (PNJL) model~\citep{Ratti:2006, Robner:2007, Ratti:2007, Fukushima:2008, Dexheimer:2010, Dexheimer:2021,Robner:2009}. In this approach, confinement is characterized by $\Phi=0$ and $\Phi\rightarrow 1$ means that deconfinement is attained. These numbers come from the definition of $\Phi$ in terms of the free energy of the system, $F$, namely, $\Phi=e^{-F/T}$: (i) $F\rightarrow\infty$ and $T$ finite lead to $\Phi=0$ (confinement), and (ii) $F$ finite and $T\rightarrow\infty$ lead to $\Phi=1$ (deconfinement).

Besides incorporating these important physics in the system, the PNJL model is not able to describe the confinement/deconfinement transition at $T=0$, since in this regime all thermodynamical quantities are reduced to the ones related to the NJL model, which does not take into account the traced Polyakov loop. A proposal to circumvent this problem was implemented in~\cite{sagun} where a dependence on the quark density was introduced in the $b_2(T)$ function of present in the Polyakov potential. Another procedure was adopted in~\cite{Mattos:2019,Mattos:2021,Mattos:prd}, in which the coupling constants of the two- and three-flavor PNJL model were made dependent on $\Phi$ as $G_s \rightarrow \mathcal{G}_s(G_s,\Phi) = G_s( 1-\Phi^2 )$, $G_V \rightarrow \mathcal{G}_V(G_V,\Phi) = G_V( 1-\Phi^2 )$, and $K\rightarrow \mathcal{K}(K,\Phi) = K( 1-\Phi^2)$. The motivation is to make these couplings vanish at the deconfinement phase, i.e. at $\Phi=1$. Here we follow the same approach in order to also allow the EQPM model encompassing deconfinement effects at $T=0$, by making the parameters $C$ and $D$ depending on $\Phi$ as,
\begin{equation}
C \rightarrow C'(C,\Phi)= C(1 - \Phi^2),
\label{cphi}
\end{equation}
and
\begin{equation}
D \rightarrow D'(D,\Phi)= D(1 - \Phi^2).
\label{dphi}
\end{equation}
Therefore, the equations of state for the new model, named here as Polyakov-Equiparticle (PEQP) model, are rewritten as
\begin{equation}
m_{i}' = m_{i0} + m_I' = m_{i0} + \frac{D'(D,\Phi)}{\rho_b^{1/3}} + C'(C,\Phi)\rho_{b}^{1/3},
\label{eq:masses-peqpm}
\end{equation}
\begin{equation}
\epsilon_{\mbox{\tiny PEQP}} = \Omega_0 - \sum_i \mu_i^* \frac{\partial\Omega_0}{\partial\mu_i^*} + \mathcal{U}_{0}(\Phi),
\label{ed-peqp}
\end{equation}
\begin{equation}
\mu_i = \mu_i^* + \frac{1}{3}\frac{\partial m_I'}{\partial \rho_b} \frac{\partial \Omega_0}{\partial m_I'}.
\label{mu-peqp}
\end{equation}
and
\begin{equation}
P_{\mbox{\tiny PEQP}} = - \Omega_{0}+\rho_b \frac{\partial m_I'}{\partial \rho_b} \frac{\partial \Omega_{0}}{\partial m_I'} - \mathcal{U}_{0}(\Phi),
\label{pres-peqp}
\end{equation}
in which we also introduced the term
\begin{equation}
\mathcal{U}_{0}(\Phi) = a_{3}T_{0}^{4}\mbox{ln}(1 - 6\Phi^{2} + 8\Phi^{3} - 3\Phi^{4}),
\label{u0}
\end{equation}
where $a_{3}$ is a dimensionless free parameter, and $T_0$ is the transition temperature for the pure gauge system. The term $\mathcal{U}_0(\Phi)$ is included in order to ensure $\Phi\ne 0$ solutions and also to limit $\Phi$ in the physical range of $0\leqslant\Phi\leqslant 1$, according to the findings of~\cite{Mattos:2019,Mattos:2021} (also as in the referred works, we consider $\Phi=\Phi^*$). We can use this feature to reconstruct the original EQP model, i.e., by taking $a_3=0$, one has $U_0=0$ and consequently $\Phi=0$, $C'=C$, and $D'=D$.

\subsection{Symmetric matter case}

Now that we have constructed the PEQP model, we are able to analyze the behavior of its thermodynamical quantities in the SU(3) system. We have defined in Sec.~\ref{stabwindow} the values of the current quarks masses according to the recent results provided by PDG~\citep{Workman:2022ynf}, but there are some other parameters that need to be defined as well, namely, $a_3$, $T_0$, $C$ and $\sqrt{D}$. The ones responsible for the gluonic sector of the model are $a_3$ and $T_0$, present in Eq.~\eqref{u0}. The latter one is fixed to be equal to $190$~MeV~\citep{Mattos:2021,Ratti:2006}, and $a_3$ was firstly used in~\cite{Dexheimer:2010}, where it is determined to reproduce lattice data and information about the phase diagram, as the authors explain there. They have obtained $a_3 = - 0.4$. However, this value is not suitable to provide solutions of $\Phi \ne 0$, when discussing symmetric matter, since their model is very different from the one presented here. Therefore, $a_3$ becomes a free parameter here and in this case, we have some freedom to test different values for it. Finally, the parameters $C$ and $\sqrt{D}$ are chosen from the stability window presented in Fig.~\ref{stability-window-symmetric-matter}. 

\begin{figure}
\includegraphics[scale=0.30]{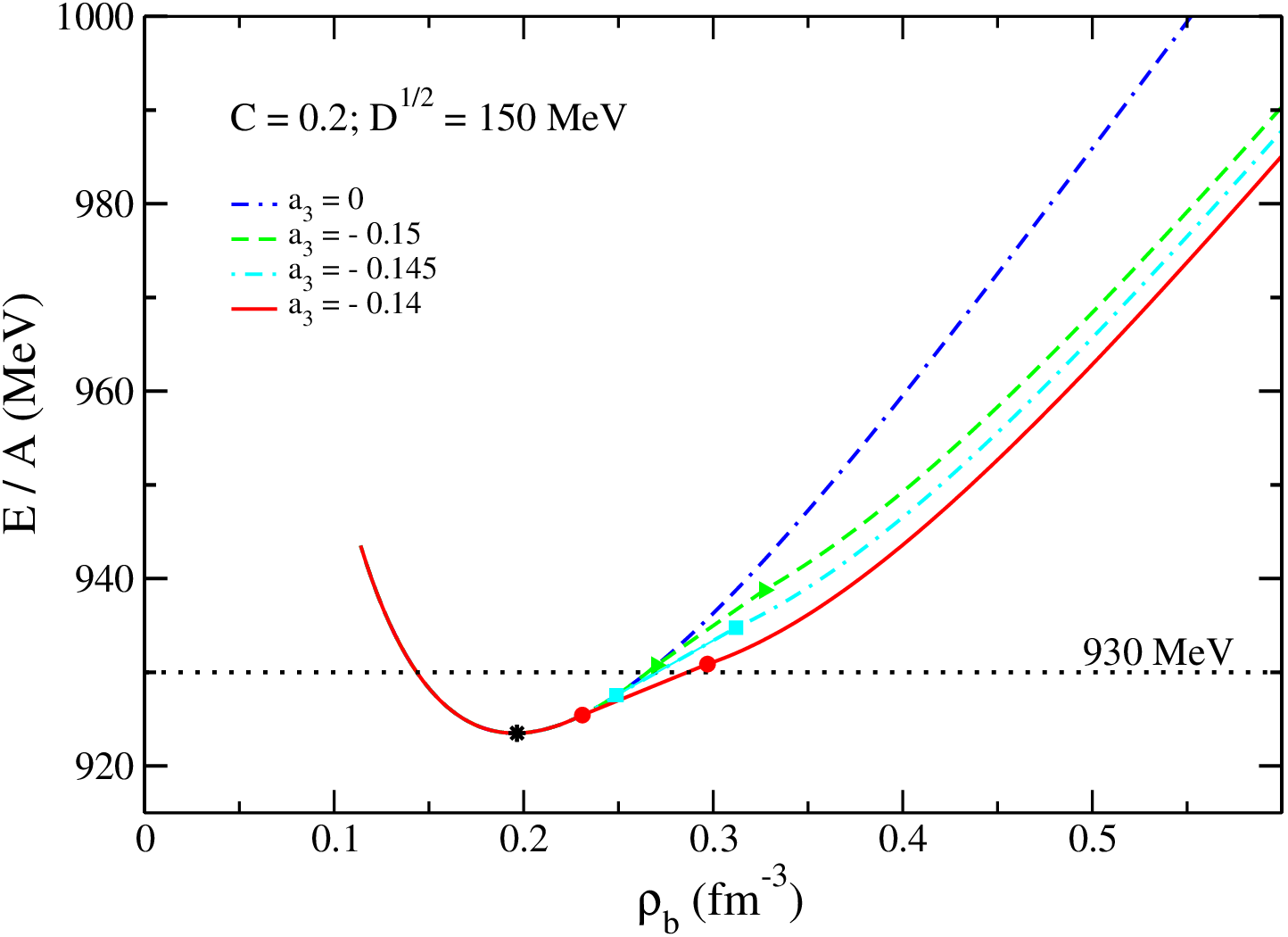}
\caption{Energy per baryon, $E/A=\epsilon/\rho_b$, as function of $\rho_b$ for EQP ($a_3=0$) and PEQP ($a_3\neq 0$) models.}
\label{sym-thermo-consist}
\end{figure}

Before performing the full analysis concerning the PEQP model, it is important to verify whether its stability windows are different from those presented in Sec.~\ref{stabwindow}. This is done, in the case of symmetric quark matter, by taking the energy per baryon with $\epsilon$ given in Eq.~\eqref{ed-peqp}, and plotting it as a function of the baryonic density for both, EQP ($a_3=0$) and PEQP ($a_3\neq 0$) models, as Fig.~\ref{sym-thermo-consist} shows. 

\begin{figure}
\includegraphics[scale=0.3]{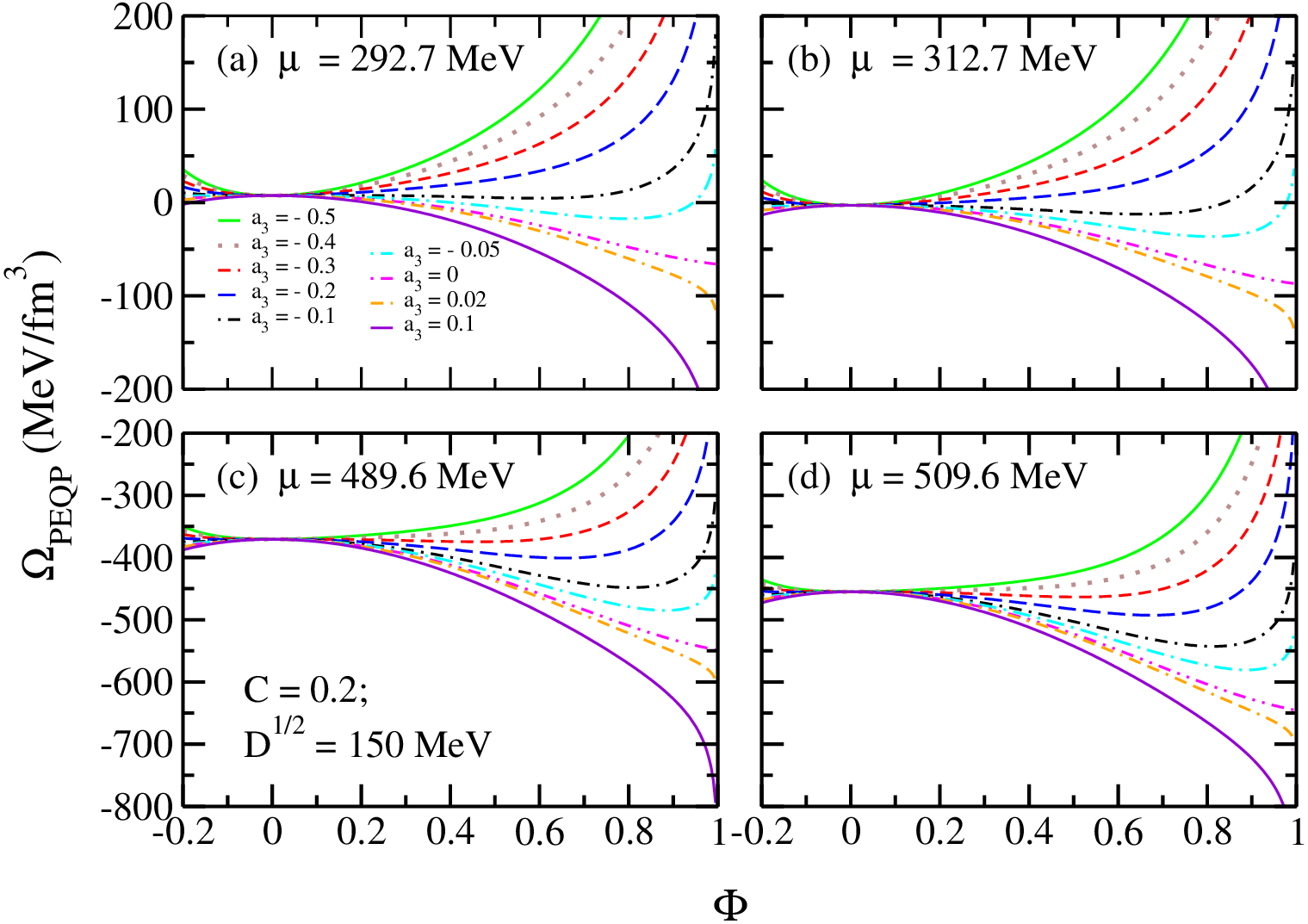}
\caption{$\Omega_{\mbox{\tiny PEQP}}$ as function of $\Phi$ for the pair $C = 0.2$, $\sqrt{D} = 150$~MeV  for different values of $\mu$, panels (a) to~(d).}
\label{Omega-Phi-first-analisys}
\end{figure}

As previously described, the minimum of the energy per baryon needs to be lower or equal to $930$~MeV in order for to SQM be stable, which can be confirmed in the figure. At this point (represented by the black dot), the pressure is zero and, therefore, the thermodynamic consistency is ensured. Notice that for the case of the PEQP model, in which $a_3\neq 0$, the minimum of $E/A$ is not modified by the emergency of $\Phi\neq 0$ solutions. In this case, it is safe to consider the same stability window shown in Fig.~\ref{stability-window-symmetric-matter} also for the PEQP model at symmetric quark matter, and then use the pairs $(C,\sqrt{D})$ that produces stable SQM. By doing so, we analyze $\Omega_{\mbox{\tiny PEQP}}=-P_{\mbox{\tiny PEQP}}$ as function of the traced Polyakov loop for the pair $C = 0.2$, $\sqrt{D} = 150$ MeV, for instance, in Fig.~\ref{Omega-Phi-first-analisys}.
For each panel, all curves were constructed by rewriting Eq.~\eqref{mu-peqp} in terms of the quarks Fermi momenta, Eq. \eqref{eq:fermi-momentum}, and the quarks densities, Eq. \eqref{eq:particle-density}, resulting in a system of three equations to determine $\rho_i$ given $\mu_u=\mu_d=\mu_s=\mu$ and $a_3$ as inputs. In this case, $\Phi$ is free to run. Notice that it is possible to obtain solutions of $\Phi\neq 0$ for some values of $a_3$ in each panel, i.e., for each chosen $\mu$.

\begin{figure}
\includegraphics[scale=0.3]{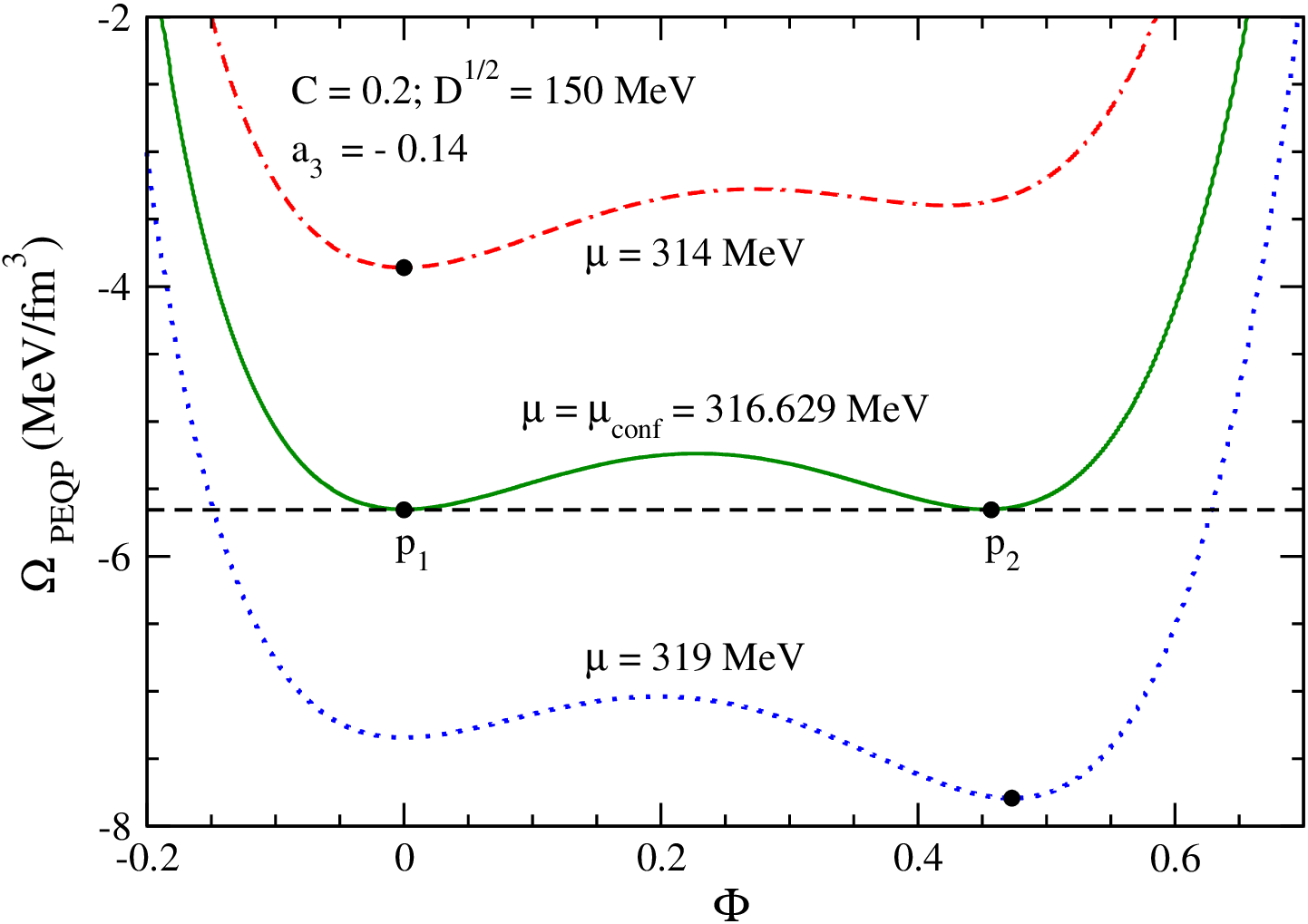}
\caption{$\Omega_{\mbox{\tiny PEQP}}$ as function of $\Phi$ for three different values of~$\mu$.}
\label{Omega-Phi-two-minima}
\end{figure}

The analysis concerning the minima in the $\Omega_{\mbox{\tiny PEQP}}\times\Phi$ curves plays an important role in the confinement/deconfinement phase transition since there is a particular value for the chemical potential, named here as $\mu_{\mbox{\tiny conf}}$, in which these curves present two minima, characterizing the transition from a confined system to a deconfined one ($\Phi$ is the order parameter related to this transition). In order to explicitly illustrate it, we plot in Fig.~\ref{Omega-Phi-two-minima} curves for $a_3=-0.14$ and different values of $\mu$.  
One can see that $\mu_{\mbox{\tiny conf}}$ = 316.629~MeV in this case. For $\mu<\mu_{\mbox{\tiny conf}}$~($\mu=314$~MeV for instance) it is only possible to obtain minima at $\Phi=0$, indicating the confined phase. On the other hand, for $\mu>\mu_{\mbox{\tiny conf}}$~($\mu=319$~MeV for instance), minima at $\Phi\neq 0$ start appearing and then deconfinement is established. Exactly at $\mu=\mu_{\mbox{\tiny conf}}$ the first order phase transition takes place, and the two minima emerge at $\Omega_{\mbox{\tiny PEQP}}\approx - 5.7$~MeV/fm$^3$ (points $p_1$ and $p_2$ in Fig.~\ref{Omega-Phi-two-minima}). The same thermodynamical analysis, performed in other contexts, can be found for instance in~\cite{yazaki}.

\begin{figure}
\includegraphics[scale=0.3]{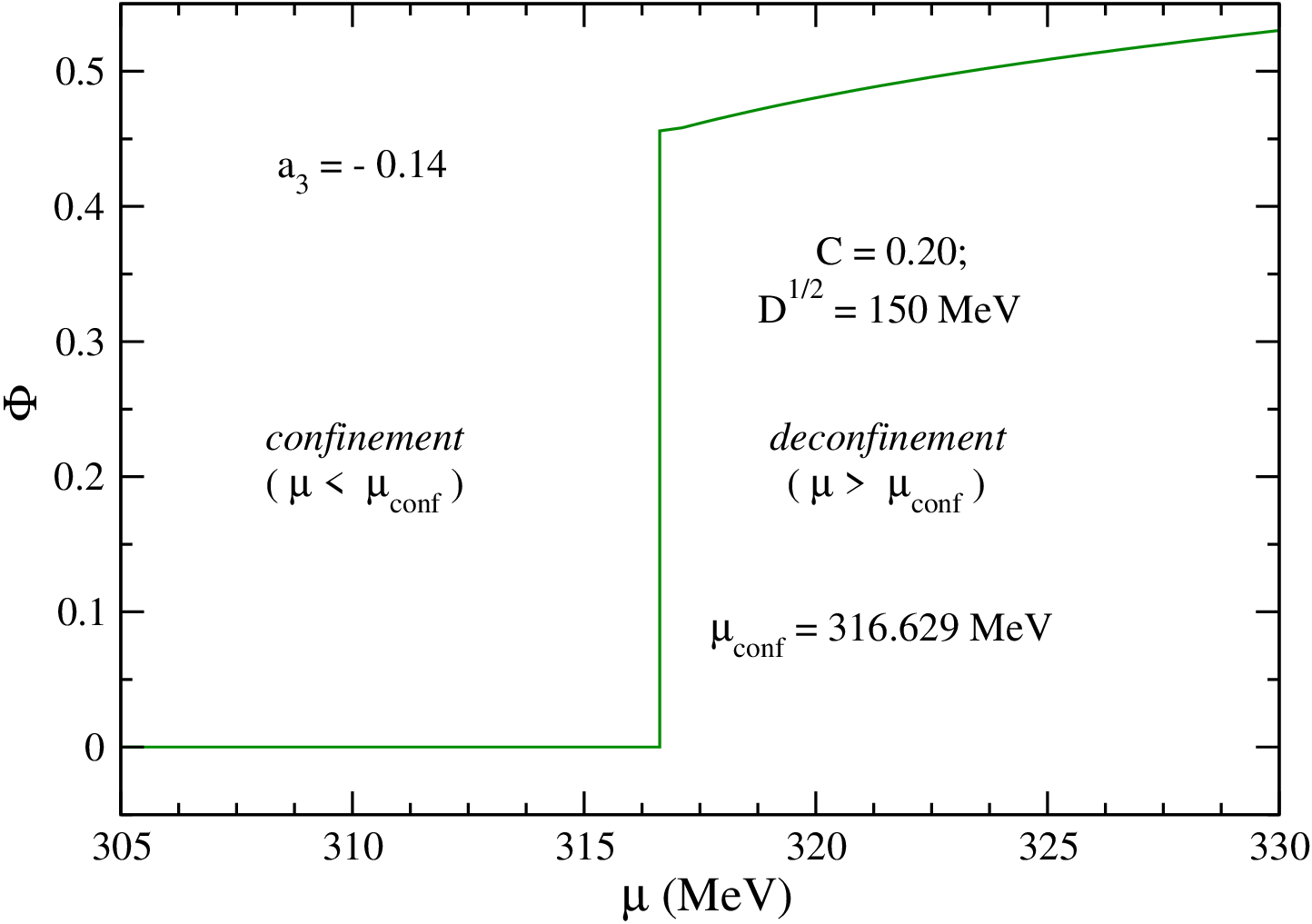}
\caption{$\Phi$ as a function of the common quark chemical potential.}
\label{Phi-Mu-analysis}
\end{figure}

\begin{figure}
\includegraphics[scale=0.3]{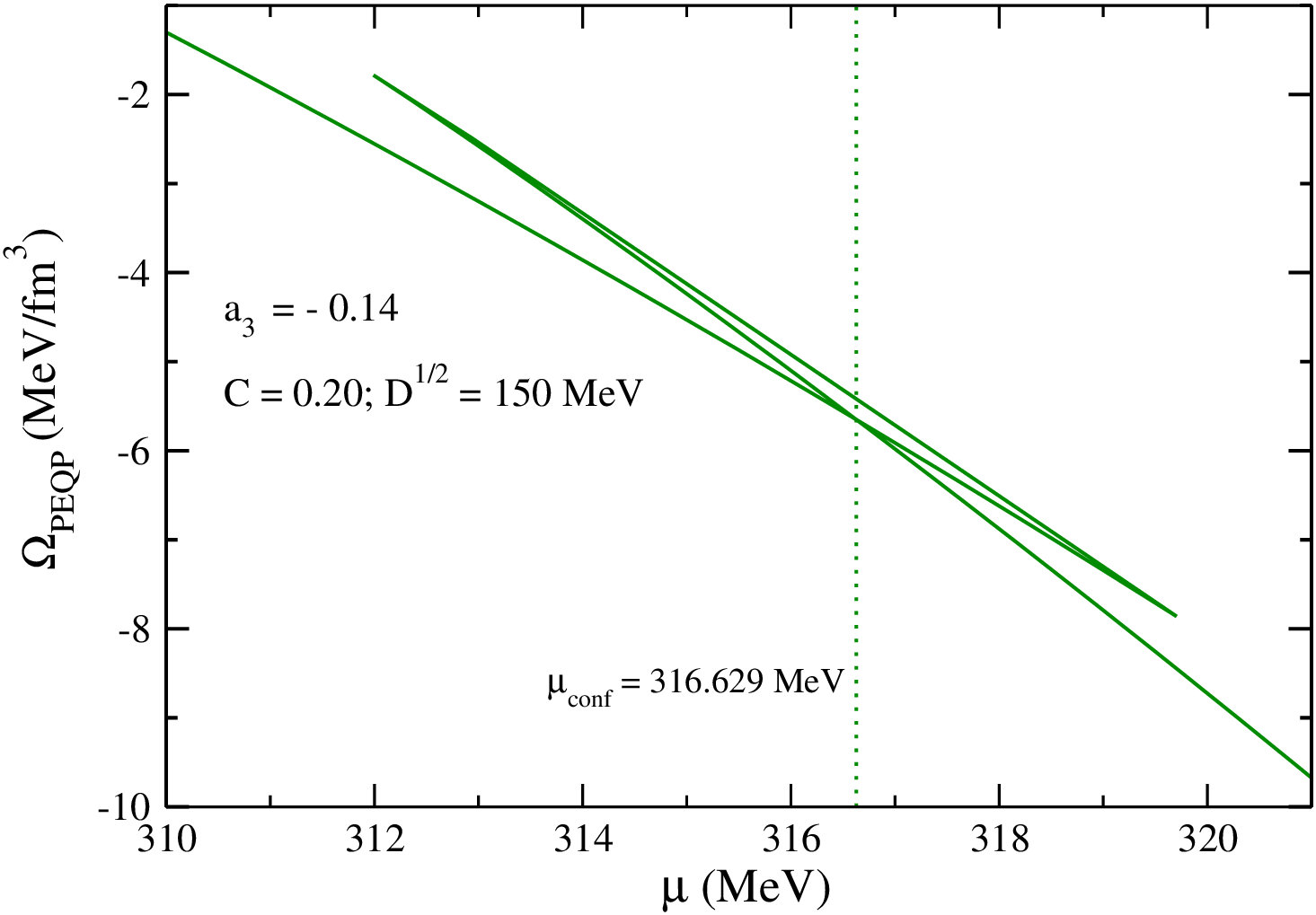}
\caption{$\Omega_{\text{PEQP}}$ as function of the common quark chemical potential.}
\label{Omega-Mu-loop}
\end{figure}

It is also possible to verify the confinement/deconfinement transition exhibited by the PEQP model from another perspective, namely, by investigating how the traced Polyakov loop depends on the chemical potential. For this purpose, we run $\mu$ and for each value used as input, we determine the respective $\Phi$ by selecting the one that minimizes $\Omega_{\mbox{\tiny PEQP}}(\mu,\Phi)=-P_{\mbox{\tiny PEQP}}(\mu,\Phi)$. In other words, we find $\Phi$ from the condition $\partial{\Omega_{\mbox{\tiny PEQP}}}/\partial{\Phi} = 0$, leading to the results displayed in Fig.~\ref{Phi-Mu-analysis}.
From the figure, we clearly identify a first-order phase transition characterized by a discontinuity in the values of the traced Polyakov loop, from $\Phi=0$ to $\Phi\ne 0$, and with the different phases being defined by the regions in which $\mu<\mu_{\mbox{\tiny conf}}$ and $\mu>\mu_{\mbox{\tiny conf}}$. These $\Phi$ solutions can now be inserted into $\Omega_{\mbox{\tiny PEQP}}(\mu,\Phi)$ in order to generate $\Omega_{\mbox{\tiny PEQP}}$ as a function of $\mu$, as depicted in Fig.~\ref{Omega-Mu-loop}.
It is worth to mention that the crossing point in the curve is located exactly at $\mu=316.629$~MeV and $\Omega_{\mbox{\tiny PEQP}}\approx -5.7$~MeV/fm$^3$. It becomes clear, therefore, that this kind of behavior presented by the grand canonical potential as a function of $\mu$ is another signature of the first-order phase transition exhibited by the model.

\subsection{Stellar matter case}

We also analyze the effect of including the traced Polyakov loop in a system composed of quarks and leptons (electrons) under charge neutrality and weak equilibrium conditions. As already mentioned in Sec.~\ref{symstellarmatter}, such restrictions must be taken into account in the description of compact stars (pure quark or hybrid stars). 

\begin{figure}
\includegraphics[scale=0.3]{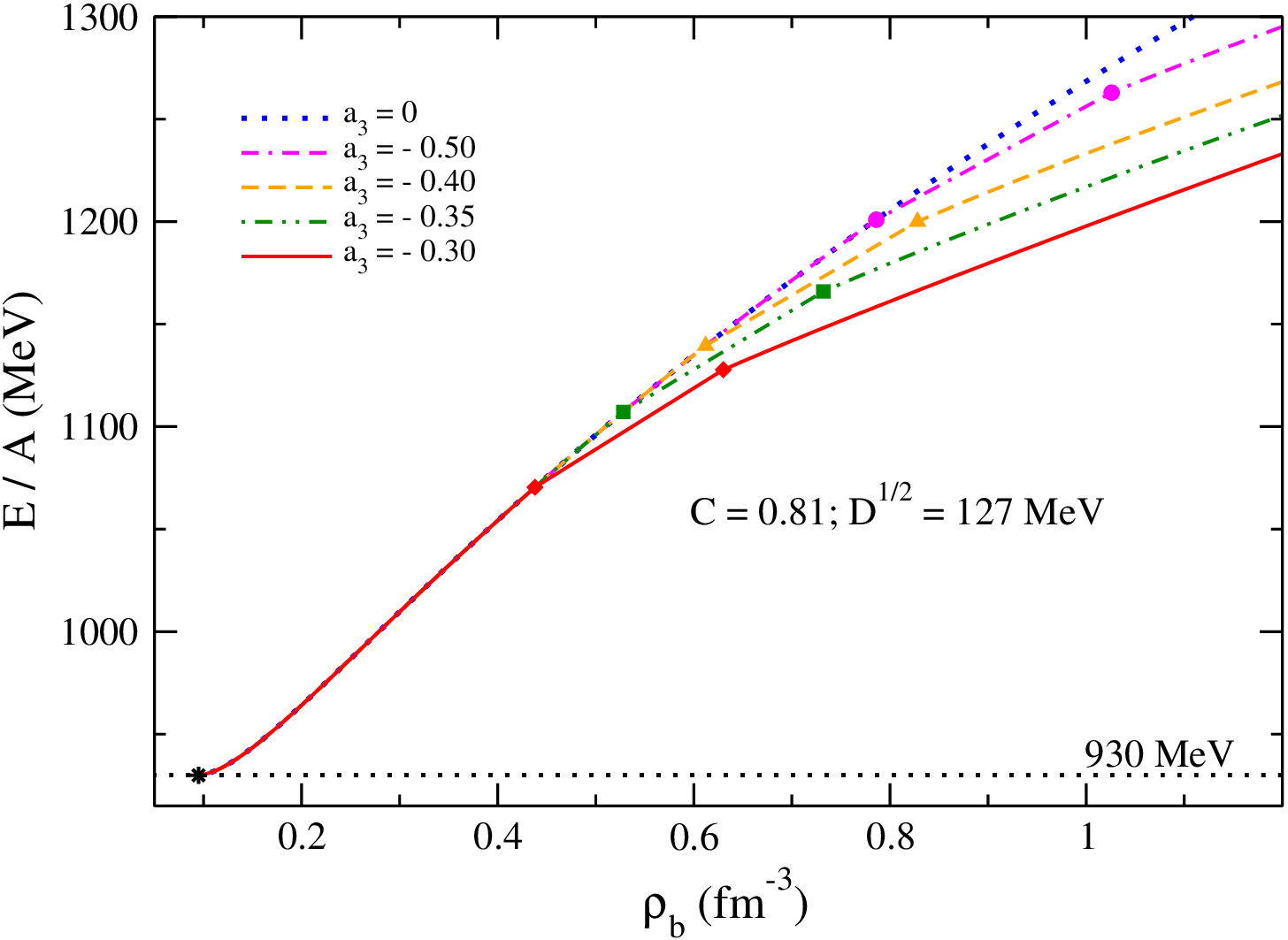}
\caption{Energy per baryon, $E/A=\varepsilon/\rho_b$, as function of $\rho_b$ for EQP ($a_3=0$) and PEQP ($a_3\neq 0$) models.}
\label{Star-DensE-Rho-loop}
\end{figure}

Because of the presence of electrons, the stability window is wider, as shown in Fig.~\ref{stability-window-stellar-matter}. This feature gives us more possibilities of choice for the parameters~$C$ and~$\sqrt{D}$. Hereafter, we take this set as being $C=0.81$ and $\sqrt{D}=127$~MeV. However, as done in the previous section, the first analysis needs to verify the stability window for the stellar matter system. This can be seen in Fig.~\ref{Star-DensE-Rho-loop}. 
In this figure we show the energy per baryon, given by
\begin{eqnarray}
\varepsilon &=& \epsilon_{\mbox{\tiny PEQP}} + \frac{\mu_e^4}{4\pi^2},
\label{det-peqpm}
\end{eqnarray}
plotted as a function of the baryonic density, of the EQP ($a_{3}=0$) and PEQP ($a_{3} \ne 0$) models. The black circle represents the minimum of the energy per baryon, which must be lower than $930$~MeV as requested for SQM to be stable. Although it is hardly visible from the curves, this value is $929.92$~MeV, and it is the same even for values of $a_{3} \ne 0$, meaning that the PEQP model maintains the minimum of the curve at the same point. This analysis allows us to rely on the stability window shown in Fig.~\ref{stability-window-stellar-matter} for the next applications of the PEQP model. 

\begin{figure}
\includegraphics[scale=0.3]{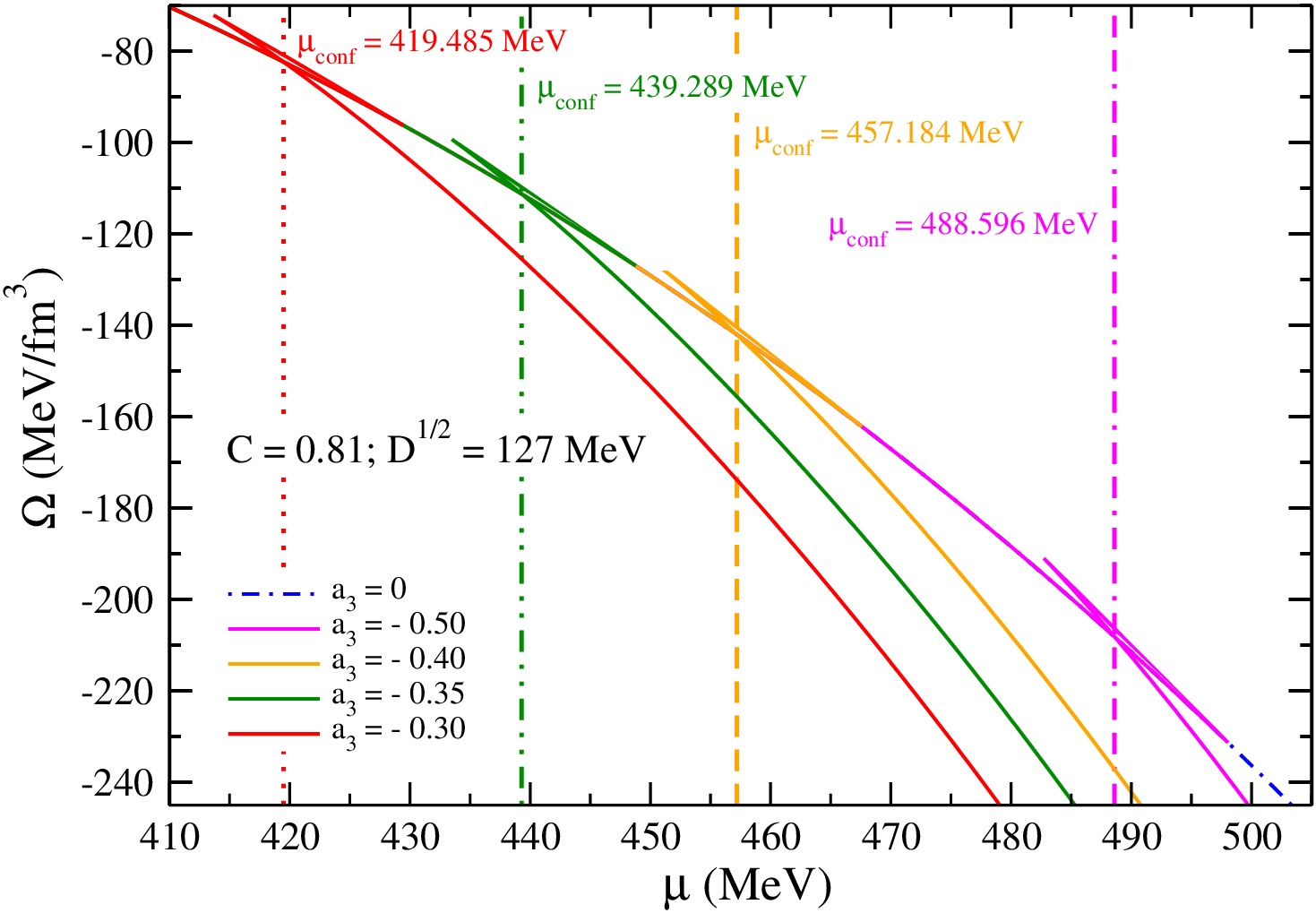}
\caption{$\Omega$ as a function of the common quark chemical potential for different values of $a_3$.}
\label{Star-Omega-Mu-loop}
\end{figure}

We now evaluate the grand canonical potential of the system through
\begin{eqnarray} 
\Omega = -p = -P_{\mbox{\tiny PEQP}} - \frac{\mu_e^4}{12\pi^2}, 
\label{presst-peqpm}
\end{eqnarray}
and display it in Fig.~\ref{Star-Omega-Mu-loop}. 
Once again, $\Omega_{\mbox{\tiny total}}$ is presented as a function of $\mu$ in order to determine $\mu_{\mbox{\tiny conf}}$. From the figure, it is possible to see the typical structure of systems that present a first-order phase transition, exactly as in the case of symmetric matter presented before. For reference, we also show the curve related to the original EQP model ($a_3=0$). For each $a_{3}$ value in the PEQP model, there is a corresponding value of $\mu_{\mbox{\tiny conf}}$. Notice that as $a_{3}$ decreases, lower values of $\Omega_{\mbox{\tiny total}}(\mu_{\mbox{\tiny conf}})$ are obtained, and for the chemical potential related to the deconfinement phase transition, the opposite happens, namely, it is higher when $a_{3}$ is lower: for $a_{3} = - 0.30$, $\mu_{\mbox{\tiny conf}}=419.485$~MeV, but when $a_{3} = - 0.50$, $\mu_{\mbox{\tiny conf}}$ increases to $488.596$~MeV.

\begin{figure}
\includegraphics[scale=0.3]{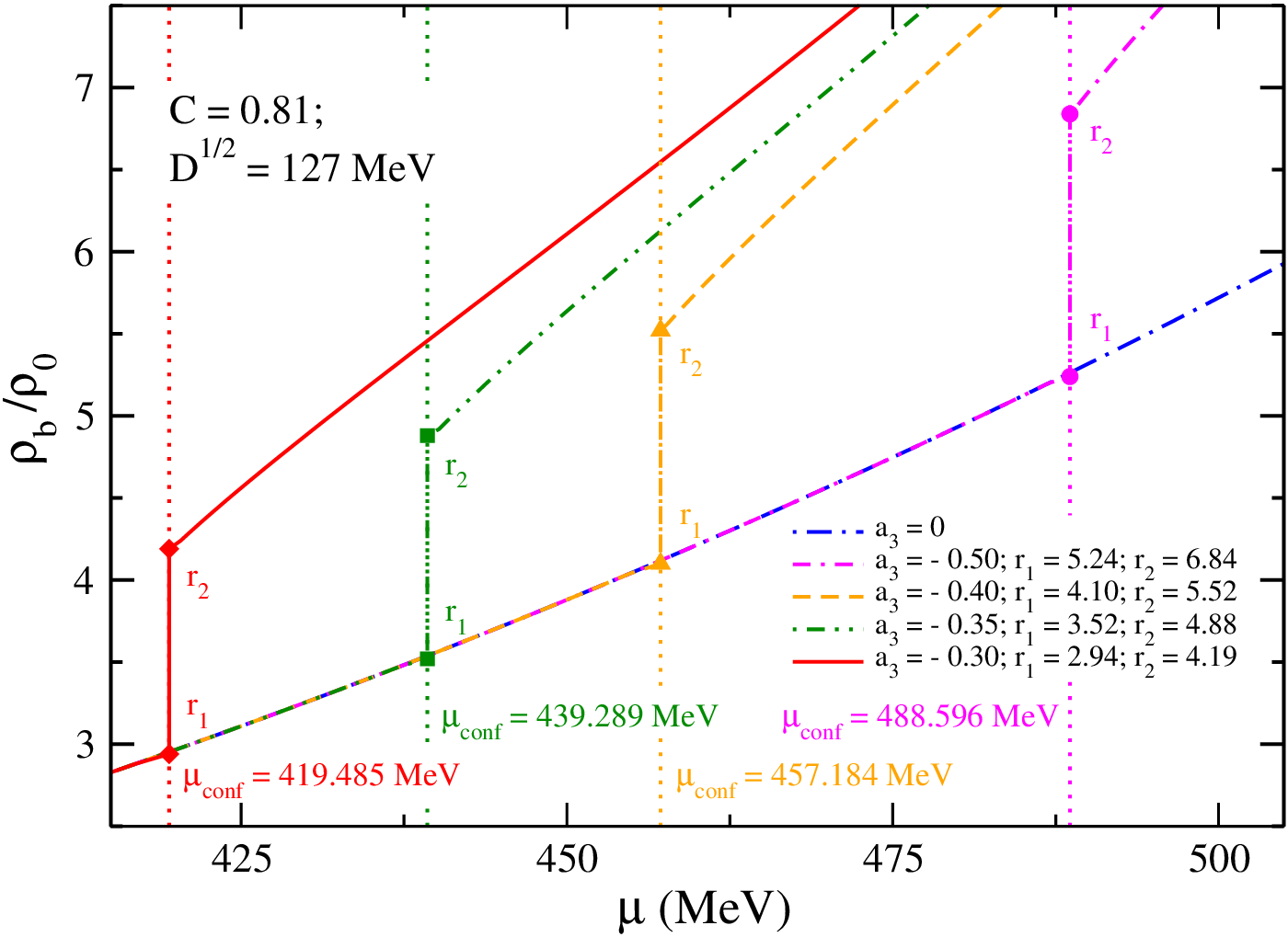}
\caption{$\rho_{b}/\rho_{0}$ as a function of the common quark chemical potential for different values of $a_3$.}
\label{Star-RhoB-Rho0-Mu-loop}
\end{figure}

Since quarks are expected to be asymptotically free at high densities, a regime found in compact stars, it is interesting to verify the values predicted by the PEQP model for the baryonic density at the phase transition point. This analysis is displayed in Fig.~\ref{Star-RhoB-Rho0-Mu-loop} where we plot the ratio of the baryonic density to the nuclear saturation density, taken here as $\rho_{0} = 0.15$~fm$^{-3}$. 
Usually, $\rho_{b}$ is expected to be at least 2 or 3 times higher than $\rho_{0}$. In the case of the PEQP model, when $a_{3} = - 0.30$, $\rho_{b}$ is $2.94$ times higher than $\rho_{0}$ when the phase transition starts, and equal to $4.19\rho_0$ at the final border of the coexistence phase. From the figure, we notice that as $a_{3}$ decreases, the ratio $\rho_b/\rho_0$ is even higher, namely, for $a_{3} = - 0.50$, $\rho_{b}$ is $5.24\rho_{0}$ at the phase transition beginning, and $6.84\rho_{0}$ at the end. For $a_{3}=0$, when the original EQP model is restored, we can see that the (blue) curve is continuous, i.e., there is no deconfined phase transitions associated. 

As a last remark concerning Figs.~\ref{Star-DensE-Rho-loop}-\ref{Star-RhoB-Rho0-Mu-loop}, we see that with a decrease of the parameter $a_3$, i.e., with increasing deviation from the EQP model ($a_3=0$), the dependencies on the density or on the chemical potential of the quarks approach the EQP model. This feature can be understood from the results displayed in Fig.~\ref{Omega-Phi-first-analisys}. Notice that the decrease of $a_3$ generates smaller $\Phi\ne 0$ solutions, i.e., the traced Polyakov loop presents values increasingly close to zero and, therefore, the system goes in the direction of recovering the original EQP model.

\begin{figure}
\includegraphics[scale=0.3]{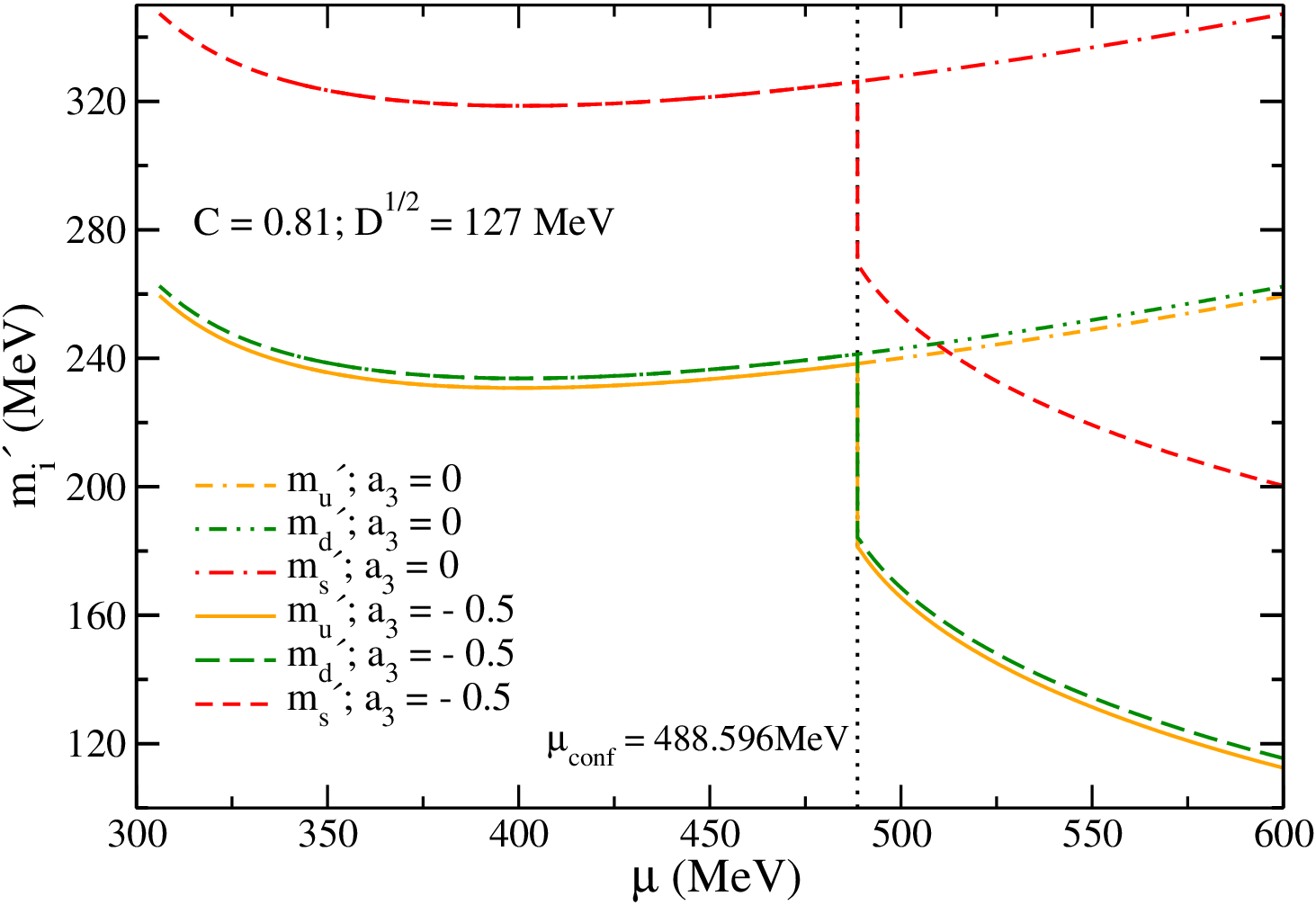}
\caption{Quark masses, calculated as in Eq.~\eqref{eq:masses-peqpm}, as function of the common chemical potential for $a_3 = - 0.50$ (PEQP model) and $a_3 = 0$ (EQP model).}
\label{Star-Quarks-Mass-Mu-loop}
\end{figure}

Another interesting investigation is about the mass of the quarks, presented in Fig.~\ref{Star-Quarks-Mass-Mu-loop}. 
Here we take a single value of $a_{3}$ and observe the behavior of each quark mass as a function of the chemical potential. The expression for the quark masses, given by Eq.~\eqref{eq:masses} for the EQP model, leads us to conclude that there will be a drop in its value and then it will slowly increase because of the latter term, as can be confirmed by the curves of the quarks $u$, $d$ and $s$ with $a_{3} = 0$ in the figure. However, for the PEQP model, the quark masses are now written as in Eq.~\eqref{eq:masses-peqpm}. The inclusion of the traced Polyakov loop results in a significant reduction of these quantities when comparing PEQP and EQP models. This is a very important result since it indicates that the new model makes the system to be in the direction of the chiral symmetry restoration, associated with the reduction of the constituent masses, specially $m_u$ and $m_d$. This effect is observed in other effective QCD quark models, such as the NJL one~\citep{njl1,njl2,njl3,njl4,njl5,njl6}. Furthermore, this feature is also verified when deconfinement phenomenology is implemented in the NJL model itself through the Polyakov loop, in the PNJL0 model developed in~\citep{Mattos:2021}.

\begin{figure}
\includegraphics[scale=0.3]{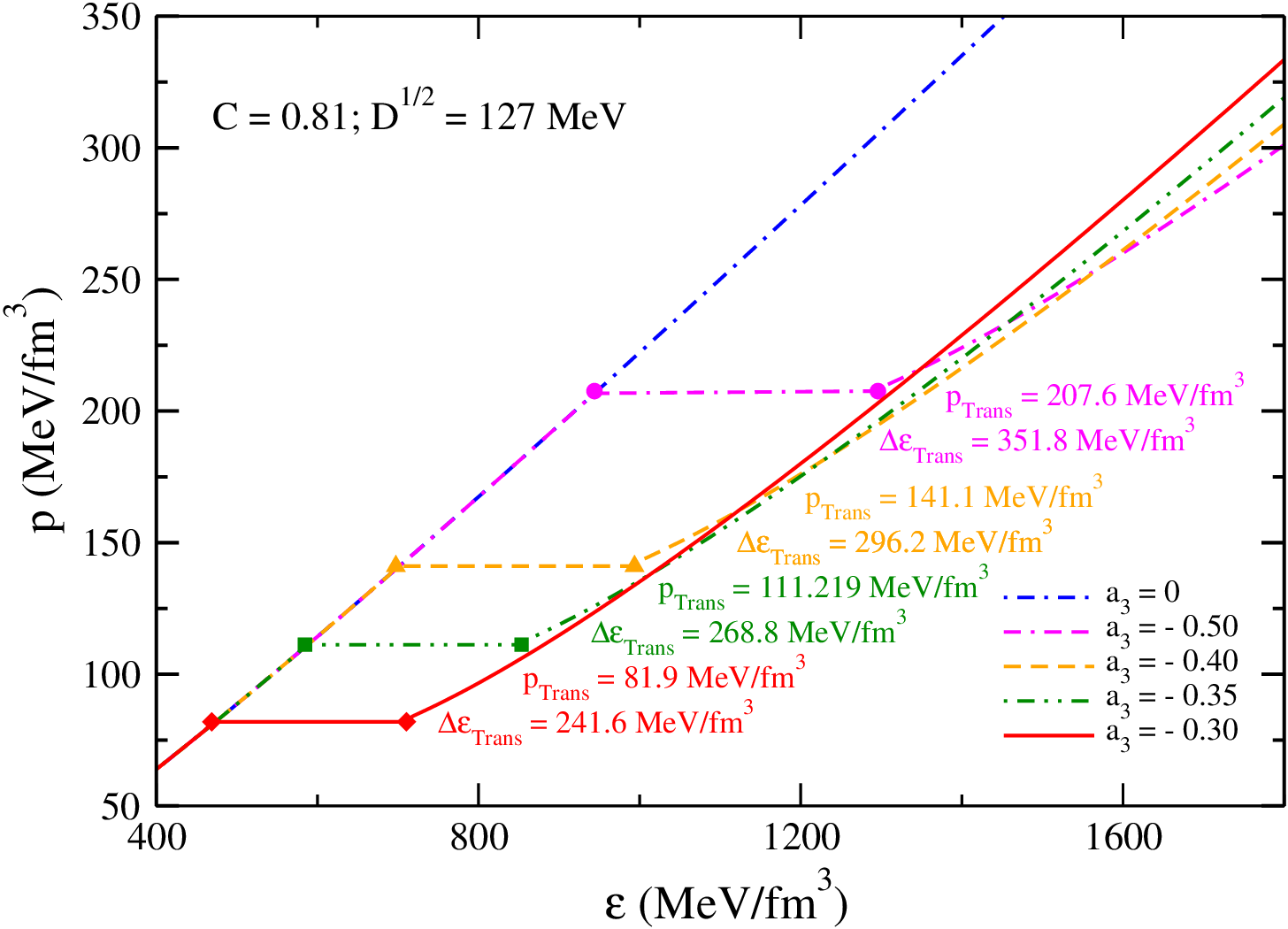}
\caption{Equation of state $p\times\varepsilon$ for different values of~$a_3$.}
\label{Star-P-E-loop}
\end{figure}

\begin{figure}
\includegraphics[scale=0.30]{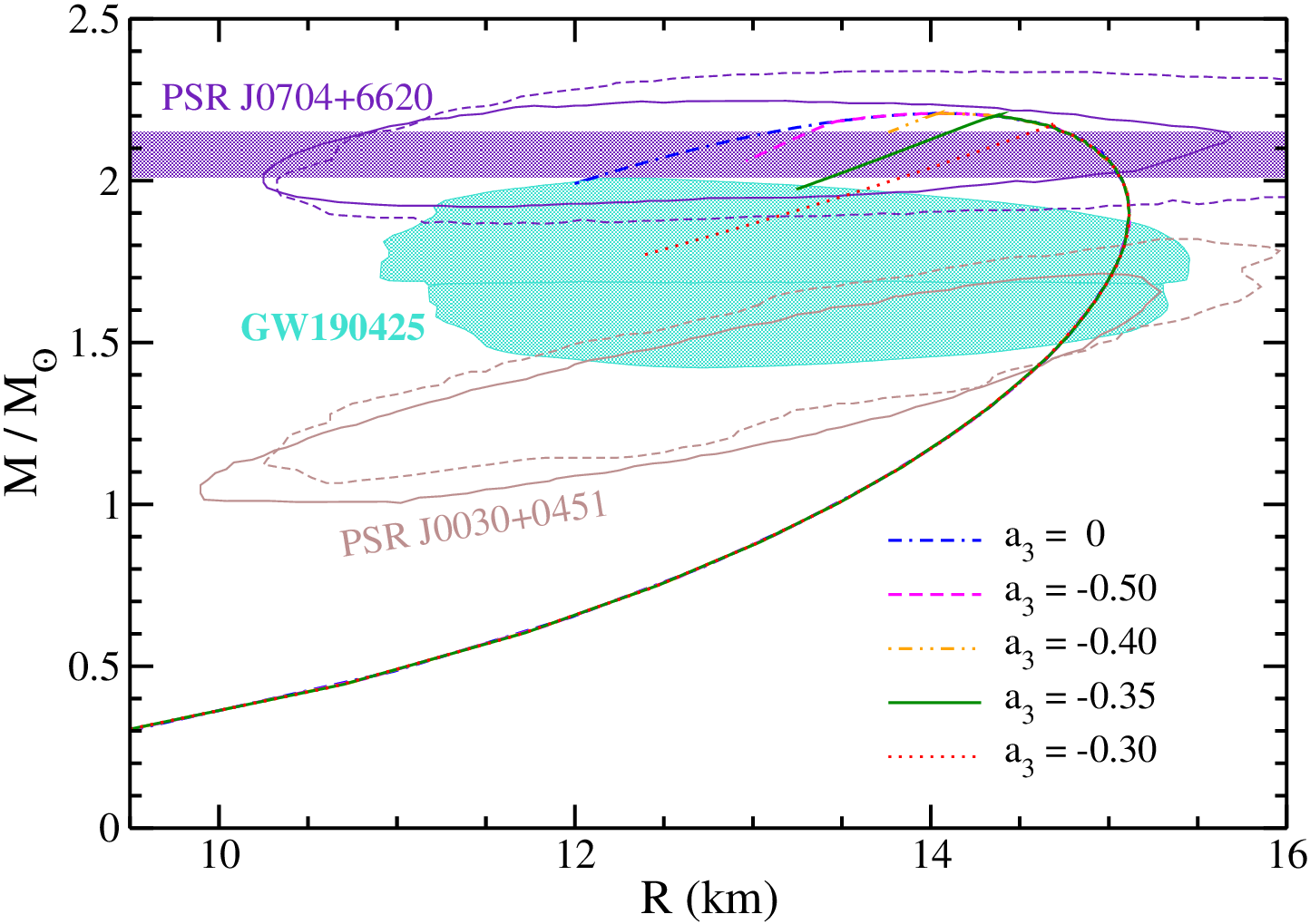}
\caption{Mass-radius diagrams constructed from the PEQP model for different values of $a_3$. The contours are related to data from the NICER mission, namely, PSR~J0030+0451~\citep{Riley_2019,Miller_2019} and PSR~J0740+6620~\citep{Riley_2021,Miller_2021}, and the GW190425 event~\citep{Abbott_2020-2}, all of them at $90\%$ credible level. The violet horizontal lines are also related to the PSR~J0740+6620 pulsar~\citep{Fonseca_2021}.}
\label{Star-M-R-loop}
\end{figure}

In Fig.~\ref{Star-P-E-loop} we show the relation between the total energy density, Eq.~\eqref{det-peqpm}, and the total pressure of the system, obtained from Eq.~\eqref{presst-peqpm}.
We notice, from the figure, another clear consequence of the confinement/deconfinement phase transition exhibited by the PEQP model, namely, the simultaneous emergence of a plateau in the pressure and a gap in the energy density. For each value of $a_{3}$, the values of the transition pressure ($p_{\mbox{\tiny Trans}}$), as well as the energy density gap at the transition point ($\Delta \varepsilon_{\mbox{\tiny Trans}}$), are also shown. From these numbers, one sees that both quantities increase when $a_{3}$ decreases. This same pattern was also observed in the study performed in~\cite{Mattos:2021} where $p_{\mbox{\tiny Trans}}$ and $\Delta \varepsilon_{\mbox{\tiny Trans}}$ were generated by a hadron-quark phase transition. The PNJL0 model (NJL model at $T=0$ with $\Phi$ included in the equations of state) was used in the quark sector of the transition. We use these curves as input to solve the Tolman-Oppenheimer-Volkoff (TOV) equations~\citep{Tolman1939,Opp1939}, given by 
\begin{align}
\frac{dp(r)}{dr} &= -\frac{[\varepsilon(r) + p(r)][m(r) + 4\pi r^3p(r)]}{r^2-2rm(r)}
\label{tov1}
\\
\frac{dm(r)}{dr} &= 4\pi r^2\varepsilon(r),
\label{tov2}
\end{align}
for which the solution is constrained to $p(0)=p_c$ (central pressure) and $m(0) = 0$, with the conditions $p(R) = 0$ and $m(R)=M$ satisfied at the star surface. $R$ is the radius of the respective quark star of mass $M$. Notice that we are using the system of units in which $G=c=1$, where $G$ is the gravitational constant, and $c$ is the speed of light. The mass-radius diagrams for the quark stars obtained from this procedure are plotted in Fig.~\ref{Star-M-R-loop}.
The hybrid star configurations are identified in this figure by the linear branches that occur for all PEQP parametrizations used in the work. In particular, notice that the emergence of all branches takes place for very massive stars, in the case of masses greater than two solar masses ($2M_\odot$). This feature is due to the confinement-deconfinement transition that happens at higher values of the pressure for all parameter sets adopted here, see Fig.~\ref{Star-P-E-loop}. If we had lower values for the transition pressure, the starting point for the hybrid stars configurations would take place at values lower than~$2M_\odot$.

The appearance of linear branches was also observed in the hadron-quark phase transitions present in the hybrid stars analyzed in~\citep{Mattos:prd} in which the quark sector was described by the PNJL0 model, which also contains the Polyakov loop in its structure. The inclusion of the Polyakov loop in both PEQP and PNJL0 models leads, in principle, to the conclusion that the hybrid stars constructed from these models are unstable since they do not satisfy the criterion of $\partial M /\partial \varepsilon > 0$. However, another kind of analysis can be performed in order to verify the stability of such star configurations. It is based on the specific response of the stars to radial oscillations~\citep{rad1,radlenzi,rad2,rad3,rad4,rad5,rad6,rad7}, that can be verified through the solution of the following coupled equations
\begin{align}
\frac{d\xi}{dr} = -\frac{1}{r}\left(3\xi + \frac{\Delta p}{\Gamma p}\right) - \frac{dp}{dr}\frac{\xi}{(p+\varepsilon)},
\end{align}
and
\begin{align}
\frac{d\Delta p}{dr} &= \xi\left[\omega^2e^{\lambda-\nu}(p+\varepsilon)r - 4\frac{dp}{dr}\right]
\nonumber\\
&+\xi\left[\left(\frac{dp}{dr}\right)^2\frac{r}{p+\epsilon}-8\pi e^\lambda(p+\varepsilon)pr\right]
\nonumber\\
&+\Delta p\left[\frac{dp}{dr}\frac{1}{p+\varepsilon}-4\pi(p+\varepsilon)re^\lambda\right],
\end{align}
with $e^\lambda=1-2m(r)/r$, $d\nu/dr=-2(dp/dr)(p+\varepsilon)^{-1}$, and $\Gamma=(1+\varepsilon/p)(dp/d\varepsilon)$. $\xi$ is the relative radial displacement, and $\Delta p$ is the pressure perturbation, both quantities time dependent as $e^{i\omega t}$ in which $\omega$ is the eigenfrequency. In the case of stellar configurations with a
first-order phase transition, as the ones we are analyzing here, it is possible to have $\omega^2>0$ after the point of maximum mass when we consider slow phase transitions~\citep{rad1,radlenzi,rad2,rad3}. That is, stable stars can be found in configurations such as those presented by the PEQP and PNJL0 models, namely, the linear branches in Fig.~\ref{Star-M-R-loop}. In this approach, the last stable star is found in the point of the mass-radius diagram where $\omega=0$. By following this method, we verify that all parametrizations of the PEQP model used to construct the mass-radius profiles present $\omega^2>0$, i.e., all of them are stable under radial oscillations when the slow phase transitions are considered. Therefore, one verifies that a particular class of twin quark stars (stars with the same
mass but different radii), namely, one of them composed by confined quarks, and the other one in which deconfined strongly interacting particles are found. Still regarding Fig.~\ref{Star-M-R-loop}, we compare our results with the recent observational astrophysical data provided by the NICER mission regarding the millisecond pulsars PSR~J0030+0451~\citep{Riley_2019,Miller_2019} and PSR~J0740+6620~\citep{Riley_2021,Miller_2021}, and with data from the gravitational wave event named GW190425~\citep{Abbott_2020-2} analyzed by LIGO and Virgo Collaboration. Additionally, we also display the PSR~J0740+6620 data extracted from~\cite{Fonseca_2021}, that corresponds to $M = 2.08\pm0.07 M_\odot$ at 68.3\% of credible level. Our findings point out to agreement between the results generated by the PEQP model and all observational data.

\section{Summary and concluding remarks}
\label{section:summary}

In this work, we proposed an improvement in a density dependent quark model with thermodynamic consistency verified~\citep{Xia:2014}, namely, the implementation of the Polyakov loop~($\Phi$) in its equations of state, effective quantity representing the gluonic dynamics of the strong interaction. This modification makes the model capable of describing, at zero temperature regime, the quark matter transition from confined to deconfined phase. We presented the main equations of the original equiparticle~(EQP) model, as well as its stability windows for recent values for current quark masses, $m_{u0} = 1.90$~MeV, $m_{d0} = 4.67$~MeV and $m_{s0} = 93.4$~MeV, all of them extracted from PDG~\citep{Workman:2022ynf}. Then, we specifically show our proposal of including $\Phi$ in the EQP model free parameters, replacing them with the following $\Phi$ dependent functions: $C'(C,\Phi)= C(1 - \Phi^2)$ and $D'(D,\Phi)= D(1 - \Phi^2)$. This requirement is done in order to ensure that interactions vanish at the deconfined phase $\Phi\sim 1$ (deconfined phase). We named the improved model as PEQP model.

The symmetric PEQP matter case was investigated. Firstly, the thermodynamic consistency was tested, allowing us to rely on the stability window provided by the original EQP model. Then, we analyzed the confinement/deconfinement phase transition through the grand-canonical thermodynamic potential, $\Omega_{\mbox{\tiny PEQP}}$, as a function of $\Phi$. It was possible to clearly identify a particular chemical potential~($\mu$) that produces two minima of these curves with the same value of $\Omega_{\mbox{\tiny PEQP}}$, an unequivocal signature of first-order phase transitions, with $\Phi$ being the order parameter in this case. Another equivalent analysis was performed by means of the plots of $\Omega_{\mbox{\tiny PEQP}}$ as a function of $\mu$. It was shown that such curves present a crossing point, a structure also used to identify first-order phase transition signatures. 

By applying charge neutrality and beta equilibrium conditions, we could also analyze the model predictions regarding stellar matter, and quark stars more specifically. Once again, the thermodynamic consistency was verified to be consistent with the original model, and its stability window for stellar matter to be reliable. Two interesting features analyzed for this case were the asymptotic freedom of quarks at high densities and the quark masses. For the first one, the results indicated that baryonic density ($\rho_b$) undergoes a discontinuity exactly at the transition point, and the deconfined phase is attained for values of $\rho_b$ in a range of around $3$ to $7$ times the nuclear matter saturation density. With regard to the quark masses, it was shown that the emerging of $\Phi\ne 0$ solutions leads these quantities to a strong reduction, indicating a trend of the system to the chiral symmetry restoration, a phenomenon associated with the constituent quark masses vanishing.

Finally, we also generated the mass-radius diagrams for the PEQP model. We verified that the decreasing of the additional free parameter of the model increases the transition pressure plateau and the gap in the energy density presented by the confinement/deconfinement phase transition. For the mass-radius profiles themselves, the PEQP model was shown to be capable of generating quark stars stable configurations in agreement with recent observational data provided by the NICER mission concerning the millisecond pulsars PSR~J0030+0451~\citep{Riley_2019,Miller_2019} and PSR~J0740+6620~\citep{Riley_2021,Miller_2021}, and by the LIGO and Virgo Collaboration regarding the gravitational wave event named GW190425~\citep{Abbott_2020-2}.

\section*{Acknowledgements}
This work is a part of the project INCT-FNA Proc. No. 464898/2014-5. It is also supported by Conselho Nacional de Desenvolvimento Cient\'ifico e Tecnol\'ogico (CNPq) under Grant No. 312410/2020-4 (O.~L.). I.~M is grateful to CNPq for financial support from the Project No. 400879/2019-0. O.~L. also acknowledge Funda\c{c}\~ao de Amparo \`a Pesquisa do Estado de S\~ao Paulo (FAPESP) under Thematic Project 2017/05660-0. O.~L. is also supported by FAPESP under Grant No. 2022/03575-3 (BPE).


\bibliographystyle{apsrev4-2}
\bibliography{references-revised}

\end{document}